\documentstyle[12pt,epsf,rotate]{article}

\newcommand{\be}{\begin{equation}}
\newcommand{\ee}{\end{equation}}

\begin{document}
\title{Numerical Simulations of the Dynamical Behavior of the SK Model}

\author{
Enzo Marinari$^{(a)}$,\\ 
Giorgio Parisi$^{(b)}$ and Davide Rossetti$^{(c)}$\\[0.5em]
$^{(a)}$  {\small  Dipartimento di Fisica and Infn, 
Universit\`a di Cagliari}\\
{\small   \ \  Via Ospedale 72, 09100 Cagliari (Italy)}\\[0.3em]
$^{(b)}$  {\small  Dipartimento di Fisica and Infn, Universit\`a di Roma}
   {\small {\em La Sapienza} }\\
{\small   \ \  P. A. Moro 5, 00185 Roma (Italy)}\\[0.3em]
$^{(c)}$  {\small  Infn, Sezione di Roma 1,}
{\small   \ \  P. A. Moro 5, 00185 Roma (Italy)}\\[0.3em]
{\small   \tt marinari@ca.infn.it giorgio.parisi@roma1.infn.it}\\
{\small   \tt davide.rossetti@roma1.infn.it}\\[0.5em]
}


\maketitle

\begin{abstract}

We study the dynamical behavior of the Sherrington Kirkpatrick model.  
Thanks to the APE supercomputer we are able to analyze large lattice 
volumes, and to investigate the low $T$ region.  We present a new and 
precise determination of the remnant magnetization and of its time 
decay exponent, of the energy time decay exponent, and we discuss 
aging phenomena in the model. We exclude validity of naive aging, and 
propose different options that fit the numerical data.

\end{abstract}  

\thispagestyle{empty}
\newpage

The study of the dynamical behavior of disordered (and complex) 
systems is receiving a large amount of attention (see for example 
\cite{DIN-UNO,FM,CUKU,NONEQU} and references therein).  On one side 
this is because the study of dynamical behaviors can shed light on a 
large realm of new and interesting phenomena (aging is one of them).  
On the other side a reliable description of the amorphous glassy state 
will surely include a crucial, non-trivial dynamical behavior.

In many cases a full analytic computation is not yet feasible and many 
analytical results are based on conjectural grounds.  Numerical 
simulations are here very useful in order to support the theoretical 
conjectures.  Unfortunately a critical issue in the dynamics is the 
dependence of the times needed to reach equilibrium on the lattice 
size.  This issue can be clarified only when both the size of the 
system is very large and the observation times are very long.  This 
problem may be bypassed by considering the behavior of the infinite 
(practically very large) system as function of the (Monte Carlo) time.  

This approach requires the study of very large systems.  Unfortunately 
in the models that have been studied better, for which a full analytic 
solution for the static behavior is available, the time needed for 
executing one dynamical step increases severely with the lattice size.  
It grows as the lattice size $N$ to the second power, $N^2$ for the 
Sherrington-Kirkpatrick model (SK) \cite{SK}, and as $N^{p}$ with 
$p\ge 3$ for the $p$-spin models.  Faster simulations can be done for 
the diluted models, but in this case we do not known the static 
solution exactly.  Moreover, given our lack of analytic control of the 
dynamic behavior, we cannot be completely sure that there are not some 
subtle differences among diluted and non diluted models.

Long range models may be classified into two different categories:

\begin{itemize}

\item
In the first class of models (taken in the infinite volume limit) 
intensive quantities like the energy $E(t)$ or the magnetization 
$m(t)$ evolve toward their equilibrium values in the limit where 
$t\to\infty$.

\item
In the second category dynamical intensive quantities do not tend to 
their equilibrium values, and truly metastable states are present.

\end{itemize}

Such difference in the dynamical behavior is strong.  It is believed 
that the models with a continuous replica symmetry breaking belong to 
the first category while models with a one step replica symmetry 
breaking belong to the second category.

The present investigation addresses this question in the case of the 
SK model \cite{PR}.  The issue is very sensitive because some 
numerical investigations have suggested that the asymptotic properties 
of the SK model are different of the equilibrium one and in particular 
that the remnant magnetization is non zero also at infinite time 
\cite{KINZEL} (that would also imply that the weak ergodicity breaking 
scenario cannot hold).  One of the results of this note is to show 
that the value of the remnant magnetization is compatible with being 
zero in the infinite volume limit and that the SK model does belong to 
the first category we have described.

In the following we will show and discuss some of our numerical 
results.  The main features that make these results relevant are that 
we have been able to work on large lattices and for low value of the 
temperature $T$.  Our use of the APE-100 supercomputer \cite{APE} has 
been crucial to allow such large scale simulations.  The fact that our 
program is truly parallelized (by dividing the spatial lattice among 
the different processors) makes possible to study very large lattices.  
Anyway the dynamics is still a sequential Metropolis one (we change 
one spin of a given system at the time).  We also 
notice that since we are dealing with the infinite range mean field 
model obtaining, as in the code we used\footnote{The code we have used 
for the numerical simulation is due to P. Paolucci and D. Rossetti, 
unpublished.}, an effective parallelization on a mesh with fixed 
connectivity is far from trivial.

In this note we will focus on the four main points we have been 
able to analyze:

\begin{enumerate}
\item  
We determine with good precision the time decay exponent of the 
magnetization.

\item  
We analyze the behavior of the remnant magnetization as a function of 
$N$.

\item  
We analyze the energy time decay exponent.

\item  
We discuss aging through the magnetization-magnetization time 
dependent correlation function.
\end{enumerate}
 
First a few details about our simulations.  We study the usual
Sherrington-Kirkpatrick model~\cite{SK,MPV}, with quenched
couplings $J$ chosen from a Gaussian distribution.  We have studied
lattices of size $N=2088$, $4032$, $8064$ and $18432$, and $T=0.25$,
$0.50$ and $0.75$ with a local Metropolis dynamics starting from a
fully magnetized state (all spins set to $+1$ state).  We study from
$100$ to $200$ realizations of the quenched couplings for each $N$
value but for the largest one, where we have order $15$ samples.  In
each realization of the quenched disorder we have followed six copies
of the system evolving independently (mainly to get a better
computational efficiency). Runs have been $10000$ lattice sweeps long.

To fit our data we have used the {\em Minuit} library from {\em 
Cernlib}, and the jack-knife approach to compute statistical errors.  
The results we present here have been obtained by using large time 
windows for the fits.

The dynamical behavior of physical quantities in systems with relevant 
que\-nch\-ed disorder is usually expected to converge to the 
equilibrium values with power laws.  The magnetization per spin on a 
lattice of size $N$ is expected to behave as

\begin{equation}
  m^{(N)}(t) \simeq m_{\infty}^{(N)} + \tilde{m}^{(N)} t^{-\delta^{(N)}}\ ,
\end{equation}
where the superscript $(N)$ (that we will omit in the following when
we can do so without creating ambiguities) indicates that the
parameters depend on the lattice size.

The previous formula needs a few comments.  In any finite system the 
residual magnetization must eventually go to zero at infinite time.  
However we can distinguish at least two time scales.  Let us consider 
for example the case of an unfrustrated system with two equal free 
energy states differing by a global spin reversal.  At first the 
system will fall in one of the two equilibrium states, and on a much 
longer time scale the system will start to oscillate among the two 
different equilibrium states.  Only in this second region of the time 
evolution the magnetization will become zero.  If the ground states of 
the system are in some sense random, we may expect that from the 
central limit theorem the residual magnetization at the end of the 
first phase of the dynamics will be proportional to $N^{-\frac12}$.  A 
similar phenomenon for large, but not too large times is expected to 
take place here.  The effective remnant magnetization is likely to be 
$N^{-\omega}$.  While in an unfrustrated system $\omega=\frac12$, the 
presence of a large number of equilibrium states suggests a smaller 
value of $\omega$ and our data are compatible with the value 
$\omega=\frac14$ found close to $T_{c}$ for the SK model.  The 
possibility $\omega=0$, i.e.  of a non decay of the remnant 
magnetization with the size, is a priori possible, but it is not the 
one preferred by our data which are well compatible with an approach 
to equilibrium for all the quantities.

\begin{figure}[hbt]
    \leavevmode
    \centering 
    \epsfxsize=300pt
    \epsffile{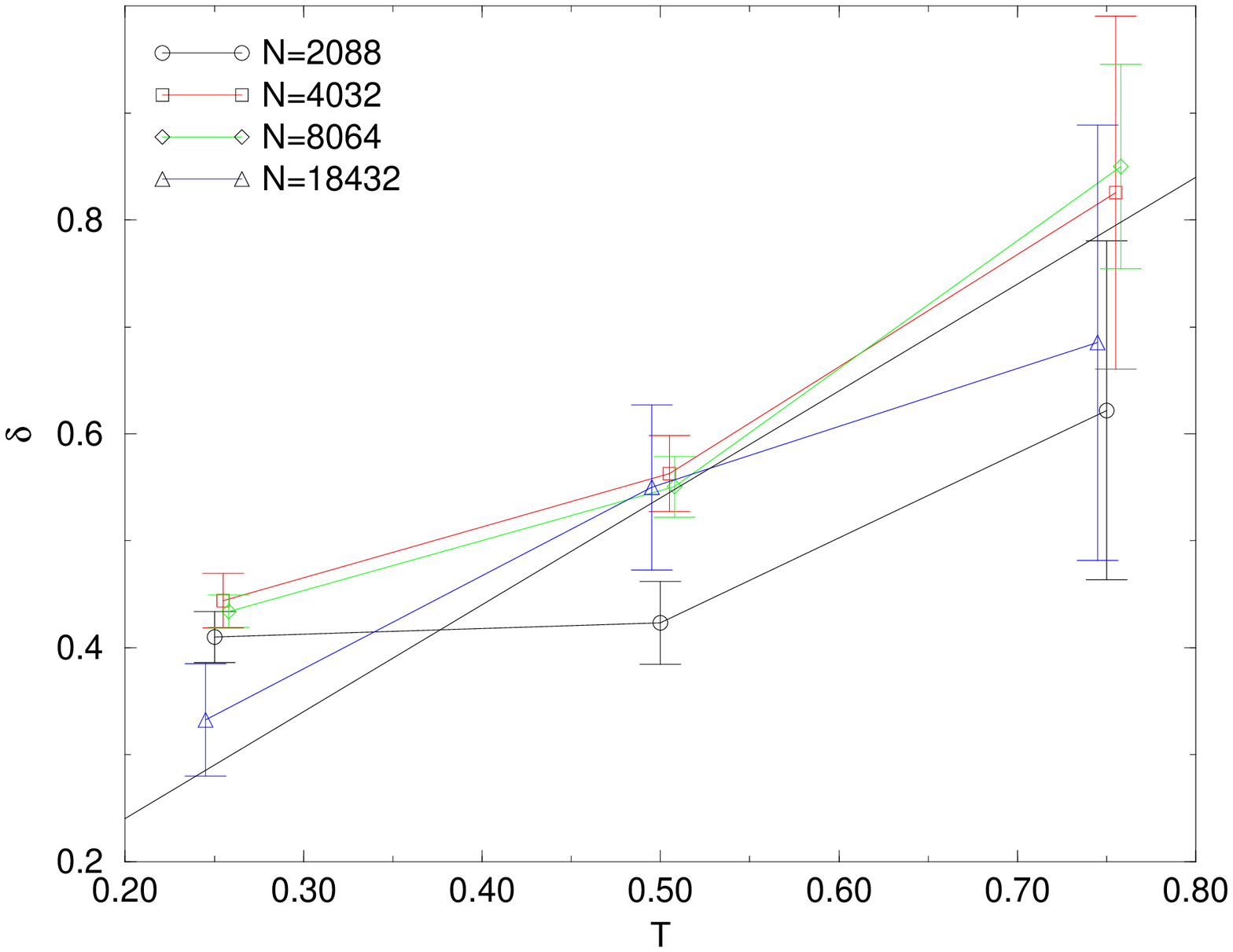}
    \protect\caption[0]{Magnetization decay exponent $\delta$ versus 
    $T$ for different $N$ values.}
    \protect\label{fig1}
\end{figure}

In figure (\ref{fig1}) we plot the exponent $\delta$ as a function of 
$T$ for the different $N$ values.  Our data are not precise enough to 
detect a clear $N$ dependence of $\delta$ (only for the smallest 
lattice, $N=2088$ one can maybe read a systematic deviation).  The 
straight line that goes through the figure is the line 
$\delta=\frac{T}{T_{c}}$: it fits very well the data, and indeed the 
fit is best on the larger lattice size, $N=18432$.  We consider this 
plot as good evidence that

\begin{equation}
  \delta(T) \simeq \frac{T}{T_{c}}\ .
\end{equation}
The exponent $\delta$ is linear in $\frac{T}{T_{c}}$, and 

\begin{equation}
  \delta(T\to T_{c}^{-}) = 1
\end{equation}
(for a discussion of the difference of this limit value  and the limit 
$\delta(T\to T_{c}^{+})$ see the second of \cite{DIN-UNO}). Our estimate is 
also compatible with the best estimate of \cite{FERRAR}, while the 
models defined on $\phi^{3}$ graphs (that is expected to have the same 
critical behavior of the SK model) seems to prefer a value 
$\delta(T) \simeq \frac{2T}{3T_{c}}$~\cite{BJMN}.
 
\begin{figure}[hbt]
   \leavevmode
   \centering 
   \epsfxsize=300pt
   \epsffile{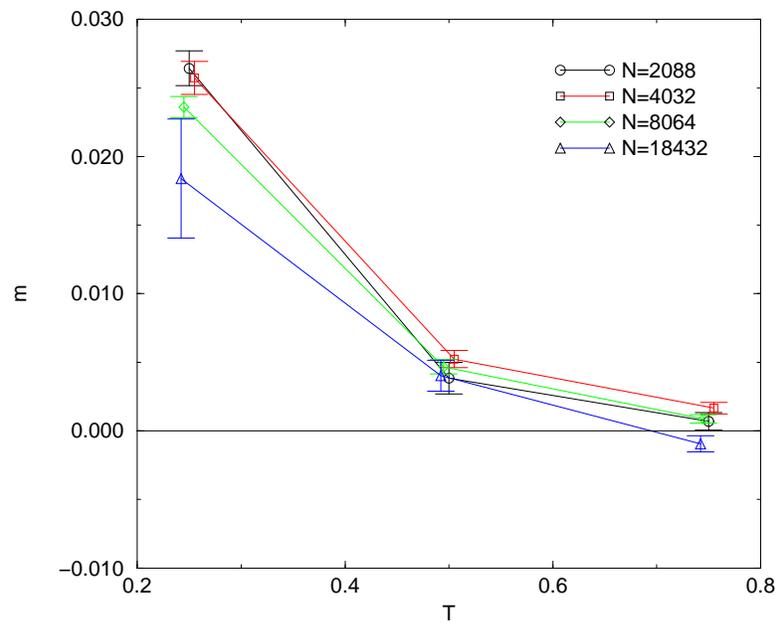}
   \protect\caption[0]{The remnant magnetization versus 
    $T$ for different $N$ values.}
   \protect\label{fig2}
\end{figure}

We plot our best estimate for the remnant magnetization as a function 
of $T$ for different $N$ values in figure (\ref{fig2}).  The $N$ 
dependence of the data is very weak.  When assuming a power decay one 
finds that the data are fully compatible with the power $\frac14$, as 
discussed in the second of \cite{DIN-UNO}.  It is clear from our data 
that a constant behavior, with a non-zero residual magnetization, 
cannot be excluded.
 
\begin{figure}[hbt]
    \leavevmode
    \centering 
    \epsfxsize=300pt
    \epsffile{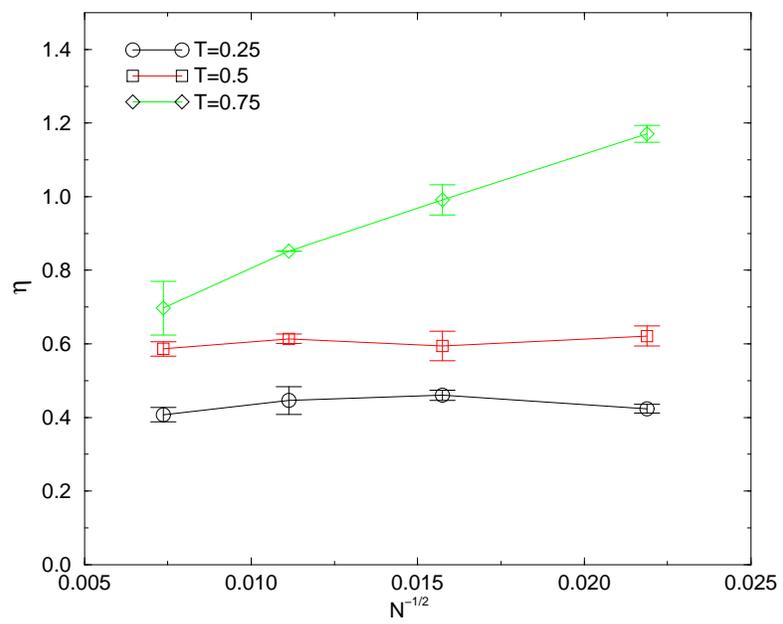}
    \protect\caption[0]{The energy time decay exponent versus 
    $N^{-1/2}$ for different $T$ values.}
    \protect\label{fig3}
\end{figure}

Our third result concerns the energy time decay exponent. In
figure (\ref{fig3}) we show it versus the inverse square root of the
volume.  At low $T$ we get an $N$-independent estimate, with a clear
dependence over $T$ (we get $.4$ at $T=.25$ and $.6$ at $T=.5$). At
$T=.75$ we seem to have a strongly $N$-dependent result. Even if in
this case it is not easy to be sure, the behavior of the power
exponent is again compatible with a linear dependence on $T$.

Our last results concern aging. We have measured the two time 
spin-spin correlation function

\begin{equation}
  C(t_{w},t_{w}+t)\equiv\frac{1}{N}
  \overline{\langle\sum_{i}\sigma_{i}(t_{w})\sigma_{i}(t_{w}+t)\rangle}\ ,
\end{equation}
always starting from random initial conditions. Our simulations were
performed at $T=.5T_c$, with ``lattice size'' $N=4096$ and $16$
independent samples with $6$ replicas each. 

The simplest scenario one can think about is naive aging, i.e. 

\begin{equation}
  C(t_{w},t_{w}+t)=f(\frac{t}{t_w})\ ,
  \protect\label{E-NAIVE-AGING}
\end{equation}
in the region where both $t$ and $t_{w}$ are large.

\begin{figure}[hbt]
  \begin{center}
    \leavevmode \centering\rotate[r]{
    \epsfxsize=220pt\epsffile{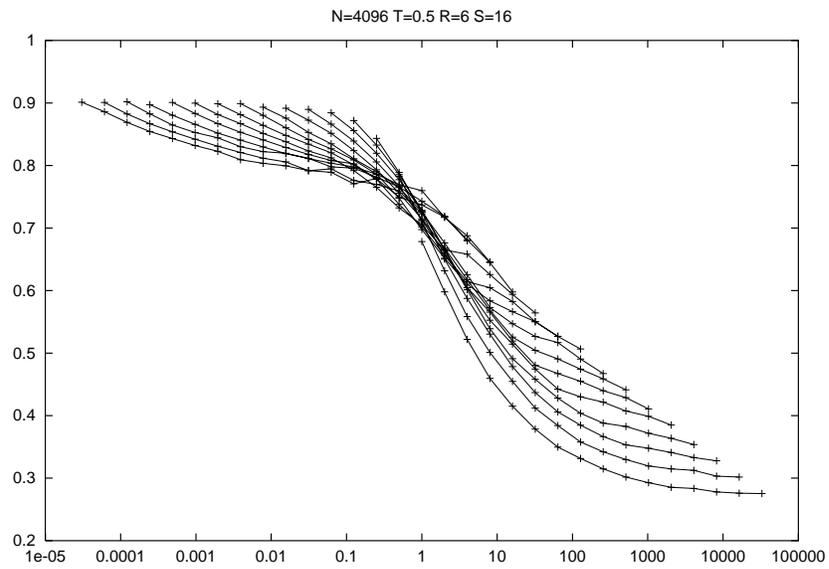} }
    \protect\caption[0]{$C(t,t_{w})$ as a function of $\frac{t}{t_w}$ 
    for different $t_{w}$ values (from 8 to 32768). $T=0.5T_{c}$.}
    \protect\label{fig6}
  \end{center}
\end{figure}

Our results for $C$ are displayed in figure (\ref{fig6}).  Although 
the data are in some very rough agreement with a naive aging behavior 
there are strong systematic corrections which obviously modify the 
functional form of (\ref{E-NAIVE-AGING}).  Naive aging is not 
satisfied in the Sherrington-Kirkpatrick model. The type of violations of 
naive aging are similar to those observed in real experiments: the 
function $C(t_{w},t_{w}+t)$ at fixed $s\equiv \frac{t}{t_w}$ decreases 
(increases) when $t_{w}$ increases for $s<1$ (for $s>1$).

In naive aging one assumes that the time scale of relaxation at 
$t_{w}$ scales as $t_{w}$ and this assumption is not correct in the 
approach of \cite{B1} and \cite{FM,CUKU}.  One can take one of many 
different attitudes.

The first possibility is that naive aging is not correct, and we 
decide to use some phenomenological corrections to it.  In this case a 
different form of scaling may be expected.  The most popular 
assumption is {\em interrupted aging} \cite{B1}.  It corresponds to 
assume that
  
\begin{equation} 
  C(t_{w},t_{w}+t) = f(\frac{t}{t_w}^{1+\mu})\ .  
\end{equation}  
The introduction of the power correction $\mu$ may help to fit the 
data at large $\frac{t}{t_{w}}$ but it does not improve the situation 
at small $\frac{t}{t_{w}}$ and spoils the agreement of aging at 
$\frac{t}{t_{w}}\simeq 1$.  It seems that this correction is not very 
useful here.  Moreover there is no theoretical justification for 
interrupted aging in this context.
  
\begin{figure}[hbt]
  \begin{center}
    \leavevmode
    \centering
    \rotate[r] {
      \epsfxsize=220pt
      \epsffile{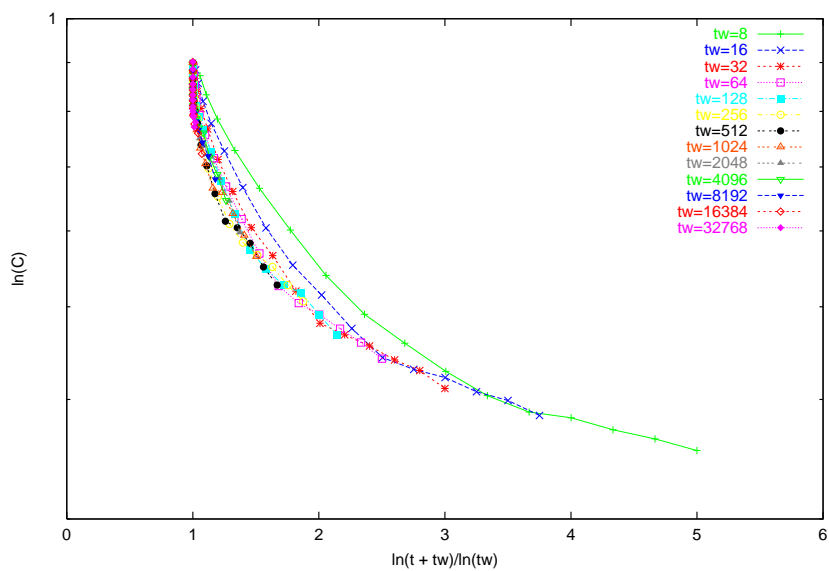} 
    } 
    \protect\caption[0]{
      $\ln C(t_{w},t_{w}+t)$ as a function of
      $\ln(t+t_{w})/\ln(t_{w})$. $T=0.5T_{c}$.
      } 
    \protect\label{fig8} 
  \end{center}
\end{figure}

Another possibility, natural in the 
contest of \cite{FM,CUKU} is to assume that
  
\begin{equation}
    C(t_{w},t_{w}+t)=f(\frac{\ln(t+t_{w})}{\ln(t_{w})})\ .
    \protect\label{THATISIT}
\end{equation} 
We have plotted our data for $C(t_{w},t_{w}+t)$ versus 
$\ln(t+t_{w})/\ln(t_{w})$ in figure (\ref{fig8}).  This change of the 
scaling law definitely improves the situation, so this is a possible 
solution to the problem.  It is interesting to note that in this 
version of the scaling form we have that

\begin{equation} 
  \lim_{t_{w} \to \infty} C(t_{w},\lambda t_{w}) 
\end{equation} 
goes to a value independent on $\lambda$, which can be therefore 
identified with $q_{EA}$.  Obviously in this case the function $f$ 
will be discontinuous when its argument is equal to one.  
We can also say that the time for reaching a given value of $C<q_{EA}$ 
is given by

\be
  t=t_{w}^{\gamma(C)}
\ee
where the function $\gamma$ is related to the function $f$.  For large 
values of $t_{w}$ the value of $C$ that can be reached after the time 
$t$ and after $\lambda t$ do coincide.  In this way the ultrametricity 
of the configurations is satisfied also in the dynamics (as it should 
be): for example it implies that if in a given time we can go at 
distance $1-C$ from a given configurations, we can arrive always at 
the same distance if we double the time.  This would be the scenario 
that is in better agreement with the existing theoretical computations 
\cite{CUKU}.

The last possibility is that naive aging is asymptotically correct, 
but there is a correction to it which vanishes as a negative power of 
time.  The rational for this choice is that we know (as we shall see 
later) that some power corrections to scaling are necessarily present 
at small $t$.
  
Here we explore if this third possibility is compatible with our 
numerical data.  We note that in the region where $t\ll t_{w}$ we 
should have

\begin{equation} 
  C(t_{w},t_{w}+t)= q_{EA}+ A t^{-\alpha}\ .
\end{equation} 
The exponent $\alpha$ is a function of $T$; it is equal to $0.5$ at 
the critical temperature and it decreases to a smaller value when $T$ 
decreases (it has been estimated to be equal to $0.36$ at zero 
temperature \cite{SOMBIS}).  As far as the value of $\alpha$ is not 
too large (and it is certainly much smaller than $0.36$ for three 
dimensional spin glasses, i.e.  close to $0.06$) the corrections to 
the scaling law cannot be easily neglected.

The simplest modification to the naive aging prediction which is
compatible with the previous equation is:

\begin{equation}
  C(t_{w},t_{w}+t)=f_{0}(\frac{t}{t_w})+
  t^{-\alpha}f_{1}(\frac{t}{t_w})\ ,
  \protect\label{CORR}
\end{equation} 
where the two functions $f_{0}(s)$ and $f_{1}(s)$ are not divergent in
the limit $s \to 0$.  In principle also the exponent $\alpha$ could be
a function of $s$, but for simplicity we assume that it is a constant.

\begin{figure}[hbt]
    \leavevmode
    \centering 
    \rotate[r]{
      \epsfxsize=220pt
      \epsffile{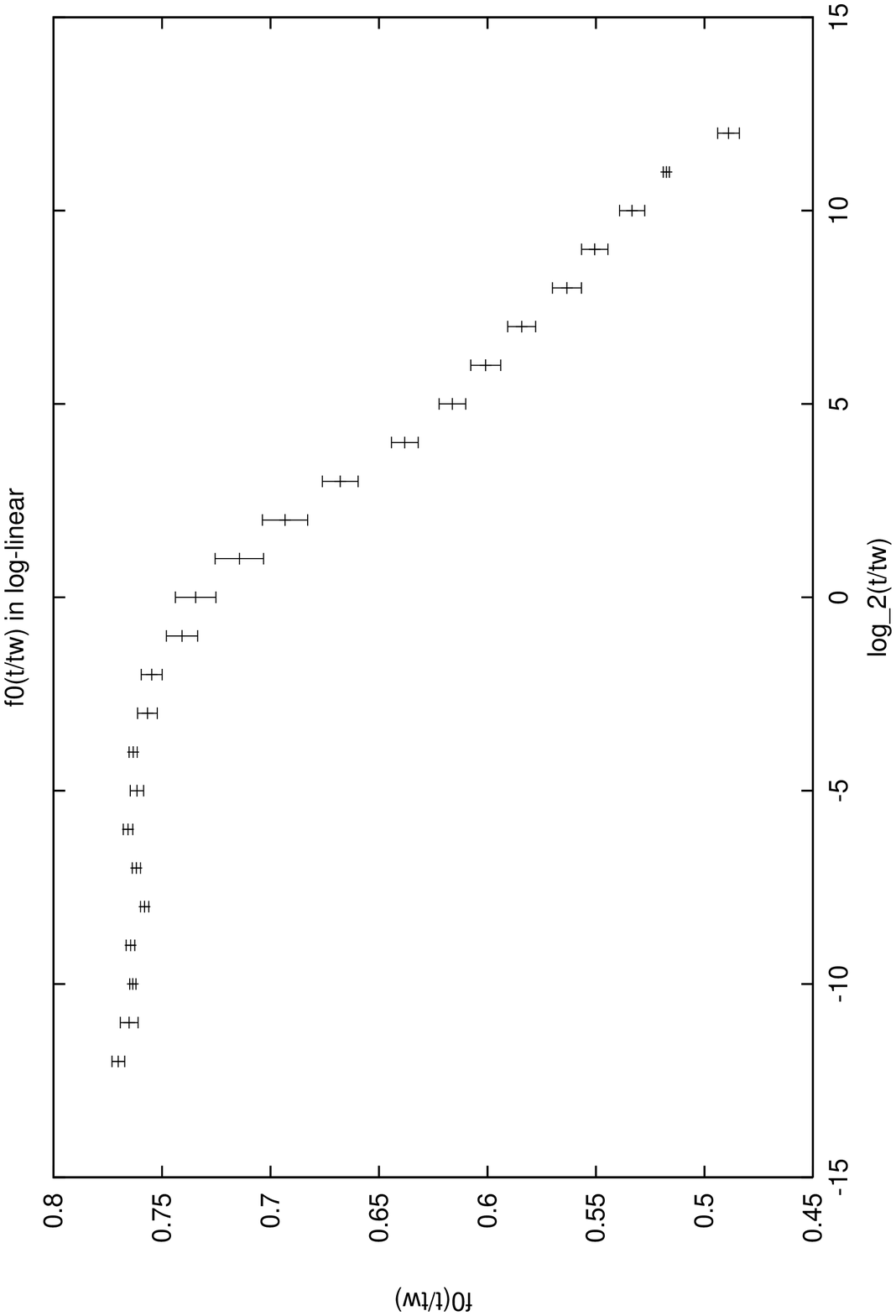} 
    } 
    \protect\caption[0]{
      Fitted $f_0$ as a function of $\frac{t}{t_w}$ for $\alpha=0.2$. 
      $T=0.5T_{c}$.
    }
    \protect\label{fig4}
\end{figure}

\begin{figure}[htb]
  \begin{center}
    \leavevmode
    \centering
    \rotate[r]{
      \epsfxsize=220pt
      \epsffile{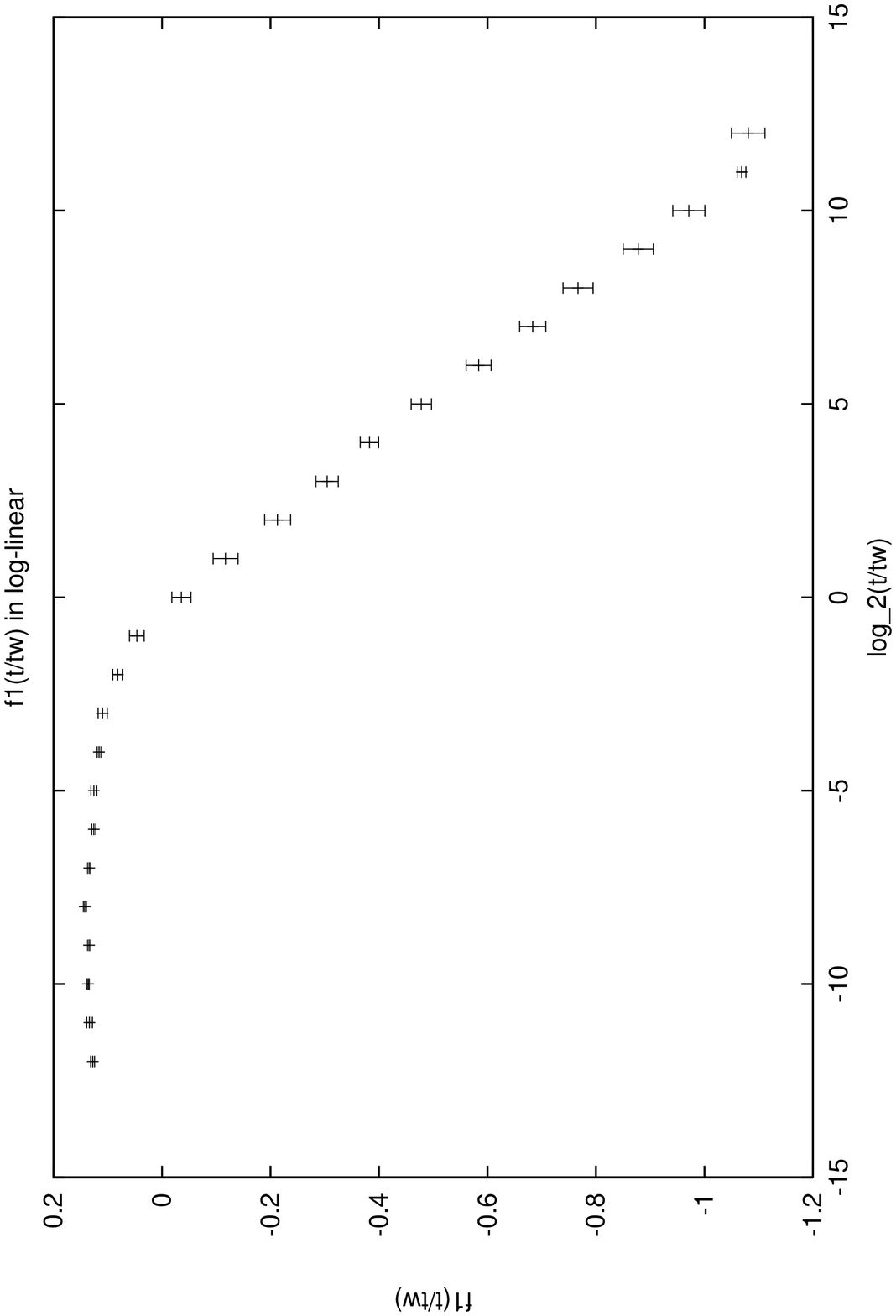} 
    }
    \protect\caption[0]{
      Fitted $f_1$ as a function of $\frac{t}{t_w}$ for $\alpha=0.2$.
       $T=0.5T_{c}$.
    }
    \protect\label{fig5}
  \end{center}
\end{figure}

We have fitted the data using equation (\ref{CORR}).  The best value
of $\alpha$ we find is $0.2$, which is a factor two smaller that the
theoretical predictions.  The origin of this discrepancy is not clear
(possible reasons are finite volume effects, a crossover behavior at
$s=0$, a strong $s$ dependence of $\alpha$).  If we stick to our best
value $\alpha=0.2$ we obtain the functions $f_{0}(s)$ and $f_{1}(s)$
displayed in figures (\ref{fig4}) and (\ref{fig5}).  In this way we
estimate a value of $q_{EA}$ close to $0.75$, which is slightly
different from the theoretical value $\simeq 0.6$.  Also in this case
the origins of this discrepancy are not clear.

\begin{figure}[htb]
  \begin{center}
    \leavevmode \centering\rotate[r]{
      \epsfxsize=220pt
      \epsffile{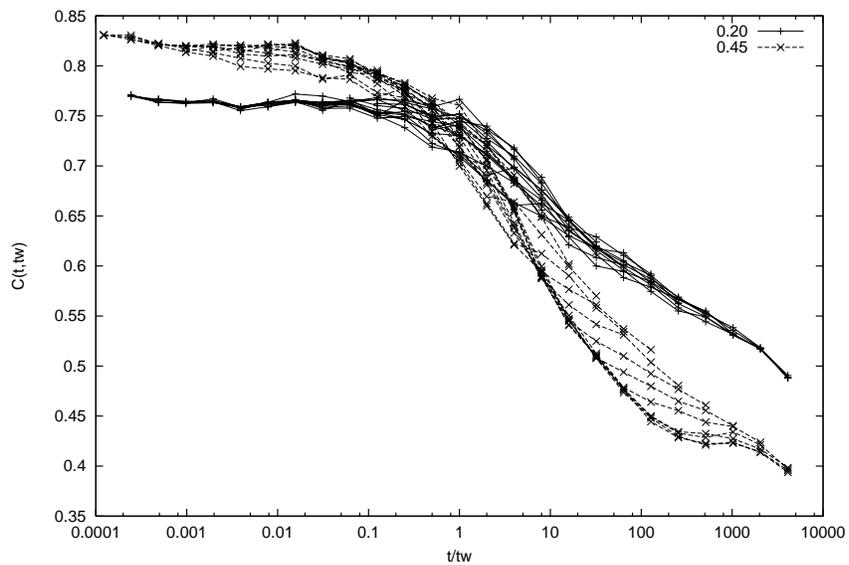}
      }
    \protect\caption[0]{
      $C(t_{w},t_{w}+t) - f_{1}(\frac{t}{t_w})t^{-\alpha}$ for 
      $\alpha=0.2$ and $\alpha=0.45$. $T=0.5T_{c}$.
      }
    \protect\label{fig7}
  \end{center}
\end{figure}

In order to exhibit the quality of our best fits we show in figure 
(\ref{fig7}) the quantity $C(t_{w},t_{w}+t) - f_{1}(\frac{t}{t_w})\ 
t^{-\alpha}$.  Here, at the expense of having introduced an extra 
function, the data seem to collapse well on a single curve.  We also 
plot the same quantity for a different value of $\alpha$, to show that 
in this case the collapse is not as good.

The main difference among the behavior of (\ref{CORR}) and the one of 
(\ref{THATISIT}) is that in the case of (\ref{CORR}) 
$\lim_{t_{w}\to\infty}C(t_{w},t_{w}\cdot (1+s))$ is a non-trivial 
function of $s$, while in the case of (\ref{THATISIT}) it does not 
depend on $s$.  The scaling (\ref{THATISIT}) is indeed in agreement 
with a picture where the barriers for reaching a value $\overline{q}$ 
of the overlap strongly depend on $\overline{q}$.

Longer runs on larger lattices will be able to improve our 
understanding of the situation. We believe that understanding details 
of the aging pattern is important, and that these results are a first 
step toward this interesting goal.

\end{document}